\documentclass[11pt]{article}
\usepackage{amssymb}
\newcommand{\bra}[1]{\mbox{$\langle \! \!$ #1 $\! \! \vert$}}

\newcommand{\cket}[1]{\mbox {#1 $\! \! \rangle$}}

\newcommand{\tx}[1]{\textrm{#1}}

\newcommand{\Fsl}[1]{#1 \! \! \! \! /}

\newcommand{\abs}[1]{\mid \! \! #1 \! \! \mid}

\newcommand{\Z}{\mathcal{Z}}
\newcommand{\Zt}{\tilde{\mathcal{Z}}} 
\newcommand{\mm}{\mathcal{M}}

\newcommand{\ie}{\emph{i.e.}}

\newcommand{\intR}{\int_{-\infty}^{+\infty}}

\newcommand{\sd}[1]{\tx{d}#1}
\author{Jesper Christiansen \thanks{email: jeschris@nbi.dk} 
\\ The Niels Bohr Institute \\ Blegdamsvej 17 
\\ DK-2100 Copenhagen \O \\ Denmark}
\title{Odd-f\/lavored QCD$_3$ and Random Matrix Theory}
\begin{document}
\maketitle
\begin{abstract}
We consider QCD$_3$ with an odd number of f\/lavors in the mesoscopic scaling region where the 
f\/ield theory f\/inite-volume partition function is equivalent to a random matrix theory 
partition function. We argue that the theory is parity invariant at the classical level if an 
odd number of masses are zero. By introducing so-called pseudo-orthogonal polynomials we are 
able to relate the kernel to the kernel of the chiral unitary ensemble with $\beta=2$ in the 
sector of topological charge $\nu=\frac{1}{2}$. We prove universality and are able to write the 
kernel in the microscopic limit in terms of f\/ield theory f\/inite-volume partition functions.
\end{abstract}

\section{Introduction}
In 1993 Shuryak and Verbaarschot realized \cite{S-V} that a certain chiral random matrix theory 
(RMT) is equivalent to the low-energy limit of the QCD partition function \cite{LS} and thus
RMT, a tool that had shown merit in such diverse areas of physics as atomic physics, solid state
physics, and nuclear physics, entered the realm of quantum chromodynamics (see \cite{genreview}
for an extensive review of the many applications of RMT and \cite{qcdreview} for a review of 
the application of RMT to QCD). There is an amount of arbitrariness in the RMT of QCD since the 
potential $V$ that enters in the RMT partition function is arbitrary. It was shown \cite{S-V} that 
when $V \sim \lambda^2$ then RMT and QCD are equivalent in the mesoscopic scaling region, also 
called the (double-)microscopic limit. Thus, if we want RMT to provide us with information 
about QCD in the microscopic limit we have to consider spectral correlators that are 
independent of $V$ in this limit. If this is the case the correlators are said to be universal.
Clearly, the question of universality is crucial if one wants to extract information about QCD 
from RMT. 

Many papers have been written about the connection between RMT and 4-dimensional QCD (QCD$_4$) 
with an arbitrary number of f\/lavors as well as 3-dimensional QCD (QCD$_3$) with an even
number of f\/lavors (see references in \cite{qcdreview}). However, for an odd number 
$N_f=2\alpha+1$ of f\/lavors the situation is less understood and only treated in 
\cite{VZ}. The reasons for this lack of balance are many. The RMT partition function is
\begin{equation} \label{Zrmt}
\Zt = \int \sd T e^{-NV(T^2)} \prod_{f=1}^{N_f} \det (T+im_f) ~,
\end{equation}
where we integrate over $N \times N$ hermitian matrices $T$ with the Haar measure. F\/irst 
of all, the ef\/fective f\/ield theory partition function, equivalent to (\ref{Zrmt})
when $V \sim T^2$, depends on whether $N$ is even or odd \cite{VZ}:
\begin{eqnarray} \label{oddpartfctNeven}
\Z^{(2\alpha+1)}_N &=& \int \tx{d} U \cosh \! \left[ \tx{Tr} (N  
 \mathcal{M} U \Gamma_5 U^{\dag}) \right] \ \ \ \ (N\tx{ even}) \\
\label{oddpartfctNodd} \Z^{(2\alpha+1)}_N &=& \int \tx{d} U \sinh \! \left[ \tx{Tr} (N 
 \mathcal{M} U \Gamma_5 U^{\dag}) \right]\ \ \ \ \ (N\tx{ odd})
\end{eqnarray} 
Both integrals are over SU$(N_f)$, $\mm$ is the mass matrix, and
$\Gamma_5=\tx{diag}(\overbrace{1, \ldots, 1}^{\alpha+1}, \overbrace{-1, \ldots, -1}^{\alpha})$. 
The $N$-dependence is most peculiar. Furthermore, the RMT spectral density $\rho(\lambda)$ obeys
the relation \cite{VZ}
\begin{equation} \label{rho}
\rho(-\lambda) ~=~ (-)^{N N_f} \rho(\lambda)
\end{equation}
which implies that for odd $N$ the spectral density is an odd function and thus not positive
def\/inite. F\/inally, the standard method for calculating spectral information in random matrix 
theories, the method of orthogonal polynomials, is not available for an odd number of f\/lavors
due to the fact that it is impossible to def\/ine orthogonal polynomials with respect to an odd
measure on an interval symmetric about the origin. 

In this paper we concentrate on the case of even $N$ and an odd number of f\/lavors 
$N_f=2\alpha+1$. In section \ref{sec: symmetries} we propose a parity transformation for an odd 
number of f\/lavors that leaves the QCD$_3$-Lagrangean invariant. This is an essential point 
since the RMT of QCD$_3$ has parity symmetry as an input, and the conclusion is that an odd 
number of masses have to be zero for the Lagrangean to be invariant (at the classical level). 
In section \ref{sec: universality} we introduce pseudo-orthogonal polynomials and calculate the
spectral correlation functions in the microscopic limit for even $N$. It turns out that for an
odd number of f\/lavors the correlators can be expressed using the kernel from the chiral 
unitary ensemble, which is equivalent to four-dimensional QCD, when we equate the number of 
chiral fermions to $\alpha$ and by analytic continuation work in the sector of topological 
charge $\nu=\frac{1}{2}$. In section $\ref{sec: finite-vol}$ we use pseudo-orthogonal 
polynomials to express the RMT kernel in terms of f\/ield theory f\/inite-volume partition 
functions. By looking at the result from \cite{PHD-GA} for the f\/inite-volume partition 
function we arrive at the conclusion that when one mass is zero, as is necessary for symmetry 
reasons, the $N$-odd f\/ield theory partition function is zero. The lesson is that the odd-$N$ 
partition function should not be regarded as def\/ining a theory in itself, but only as a 
function that is necessary in order to extract spectral information in the even-$N$ theory. 
This is consistent with the fact that the spectral density does not make sense when $N$ is odd
due to (\ref{rho}).

\section{Discrete Symmetries of QCD$_3$} \label{sec: symmetries}
Our starting point is the QCD$_3$-Lagrangean with $N_f$ f\/lavors $\psi_f$:
\begin{equation} \label{L_QCD}
\mathcal{L} ~=~ - \frac{1}{4}\tx{Tr}F^2 + 
\sum_{f=1}^{N_f} \bar{\psi}_f \Fsl{D} \psi_f +
\sum_{f=1}^{N_f} m_f \bar{\psi}_f \psi_f  ~,
\end{equation}
where $F$ is the gluonic f\/ield strength and $m_f$ are the eigenvalues of the mass matrix.
In three dimensions we do not have chirality at our disposal because 
$\gamma^0 \gamma^1 \gamma^2 \propto 1$. The lowest dimensional representation of the 
$\gamma$-matrices is the Pauli-matrices and the corresponding f\/ields are two-spinors. 
This has the disadvantage that the mass terms are parity 
$P:(x^0,x^1,x^2) \mapsto (x^0,-x^1,x^2)$ \footnote{Notice that 
$(x^0,x^1,x^2) \mapsto (x^0,-x^1,-x^2)$ is a rotation. We choose (somewhat arbitrarily) to let 
$P$ denote inversion around the 1-axis.} odd. If all masses are zero then $\mathcal{L}$ is 
invariant under $P$ since the Dirac-term in the Lagrangean respects the parity transformation, 
but in general the masses are non-zero and $\mathcal{L}$ is not invariant. For reasons that 
will become apparent later we distinguish between an even and an odd number of f\/lavors. 

In \cite{Jack-Temp, Pisarski, Redlich} it is argued that for an even number $N_f=2\alpha$ of 
f\/lavors we can achieve parity invariance by using a four-dimensional representation of 
the $\gamma$-matrices. The representation employed (Minkowski space) is
\begin{equation} \label{representation}
\gamma^0 ~=~ \left( \begin{array}{cccc}
\sigma_3 & 0 \\ 0 & -\sigma_3 \end{array} \right)
\ , \quad
\gamma^1 ~=~ \left( \begin{array}{cccc}
i \sigma_1 & 0 \\ 0 & -i \sigma_1 \end{array} \right)
\ , \quad
\gamma^2 ~=~ \left( \begin{array}{cccc}
i \sigma_2 & 0 \\ 0 & -i \sigma_2 \end{array} \right)
~.
\end{equation}   
We see that for a four-spinor $\psi=(\phi \ \ \chi)^{\tx{T}}$ the mass term is
\begin{eqnarray} \label{Lmass}
\mathcal{L}_{\tx{\tiny{mass}}} ~=~ m \bar{\psi} \psi 
                               ~=~ m \psi^{\dag} \gamma^0 \psi
                               ~=~ m \phi^{\dag} \sigma_3 \phi 
                                -m \chi^{\dag} \sigma_3 \chi ~.
\end{eqnarray}
The last expression is identical to $m \bar{\phi} \phi  -m \bar{\chi} \chi$ in a 
two-dimensional representation of the $\gamma$-matrices with $\gamma_0=\sigma_3$. If the mass
matrix $\mathcal{M}$ is chosen to be
\begin{equation} \label{massmatrix}
\mathcal{M}~=~
\tx{diag}(m_1, \ldots , m_{\alpha}, -m_1, \ldots, -m_{\alpha}) 
\end{equation}
it follows that the Lagrangean (\ref{L_QCD}) is invariant under parity when the transformation
is def\/ined as
\begin{equation} \label{parity}
\begin{array}{lll}
P:~ \psi_i  &\mapsto& \sigma_1 \psi_{\alpha+i} \\
P:~ \psi_{\alpha+i}  &\mapsto& \sigma_1 \psi_i  
\end{array} \quad , \ i=1,\ldots,\alpha  ~.
\end{equation} 

For an odd number $N_f=2\alpha+1$ of f\/lavors we write the Lagrangean as
\begin{eqnarray}
\mathcal{L} ~=~ - \frac{1}{4}\tx{Tr}F^2 + 
\mathcal{L}^{(2 \alpha)}_{\tx{\tiny{Dirac}}} \! \! \! \! &+& \! \! \! \! 
\mathcal{L}^{(2 \alpha)}_{\tx{\tiny{Dirac}}} + 
\mathcal{L}^{(1)}_{\tx{\tiny{mass}}} +
\mathcal{L}^{(1)}_{\tx{\tiny{mass}}} \\
\mathcal{L}^{(2 \alpha)}_{\tx{\tiny{Dirac}}} ~=~ 
\sum_{f=1}^{2\alpha} \bar{\psi}_f \Fsl{D} \psi_f  &,& 
\mathcal{L}^{(1)}_{\tx{\tiny{Dirac}}} ~=~ \bar{\psi} \Fsl{D} \psi \\
\mathcal{L}^{(2 \alpha)}_{\tx{\tiny{mass}}} ~=~~ 
\sum_{f=1}^{2 \alpha} m_f \bar{\psi}_f \psi_f & , & 
\mathcal{L}^{(1)}_{\tx{\tiny{mass}}}~=~ m \bar{\psi} \psi ~.
\end{eqnarray}
The $\gamma$-matrices are represented by the Pauli-matrices:
\begin{equation} \label{lonegamma}
\gamma{}^0~=~\sigma_3 \quad , \quad
\gamma{}^1~=~i \sigma_1 \quad , \quad
\gamma{}^2~=~i \sigma_2 ~.
\end{equation}
Above we used the grouping of the spinors $\psi_1 , \ldots , \psi_{2\alpha}$ into four-spinors 
and the $4 \times 4$-matrices (\ref{representation}) as $\gamma$-matrices to obtain a parity 
invariant system when assigning masses of opposite sign to the constituents of the 
four-spinors. Here we use that if $\{ \gamma_{\mu} \}$ is a representation of the Dirac algebra
then $\{ -\gamma_{\mu} \}$ is an inequivalent representation of the algebra \cite{Sohnius}. 
This allows us to use negative masses also for an odd number of f\/lavors, where we cannot
group the spinors as in (\ref{Lmass}). As far as $\mathcal{L}^{(2 \alpha)}_{\tx{\tiny{mass}}}$ 
and $\mathcal{L}^{(2 \alpha)}_{\tx{\tiny{Dirac}}}$ are concerned, we use the transformation 
def\/ined in (\ref{parity}). For the remaining two-spinor $\psi$ we employ the 
transformation
\begin{equation}
P: ~\psi  ~\mapsto~ \sigma_1 \psi  ~.
\end{equation}
Since a mass term changes sign under this transformation the overall symmetry is retained 
only when $m=0$. This is also the case if we had used the representation (\ref{lonegamma}) with 
opposite signs on all the matrices. Provided that we want to study an invariant Lagrangean we 
are therefore conf\/ined to consider an \emph{odd} number of massless fermions for odd 
$N_f$. Notice, that if we want to have an optimal amount of discrete symmetry in the theory, 
with as many non-zero mass terms as possible, there is no alternative to the choice 
\begin{equation}
\mathcal{M}~=~\tx{diag}(m,m_1,\ldots,m_{\alpha},-m_1,\ldots,-m_{\alpha})
\end{equation}
(with $m=0$) of the mass matrix. We have an odd number of two-spinors that have to transform 
among each other. This means that the transformed f\/ields are permutations of the original 
f\/ields (multiplied by a Pauli matrix such that mass terms change signs). Unless the f\/ields 
swap positions two and two their masses have to be zero in order to have invariance, and when 
two f\/ields interchange positions the $\pm$ assignment of the masses make sure that the mass 
terms are invariant. This automatically leaves one f\/ield that has to transform into itself, 
thus giving rise to a symmetry breaking mass term unless $m=0$.

Having discussed the choice of mass matrix we now turn to the RMT of odd-f\/lavored QCD$_3$.

\section{Universality of Odd-f\/lavored QCD$_3$ in the Microscopic Limit} 
\label{sec: universality}
The QCD$_3$ RMT partition function for an odd number of f\/lavors in the eigenvalue 
representation
of the unitary ensemble is \cite{VZ} 
\begin{equation}
\Zt(\mm)~=~\intR \prod_{k=1}^N \sd \lambda_k (\lambda_k +im) 
\prod_{f=1}^{\alpha} (\lambda_k^2+m_f^2) e^{-NV(\lambda_k^2)} \Delta(\{\lambda_i\})^2
\end{equation}
where 
\begin{equation} \label{Vandermonde}
\Delta(\{\lambda_i\}) ~=~ \prod_{i<j}^N (\lambda_i-\lambda_j)
\end{equation}
is the Vandermonde determinant. Notice, that it is precisely the choice $m=0$, which we now
make, that makes the partition function real. With
\begin{equation} \label{Npoint}
\rho_N(\lambda_1,\ldots,\lambda_N)~=~ \frac{1}{\Zt(\mm)}
\prod_{k=1}^N \lambda_k \prod_{f=1}^{\alpha} (\lambda_k^2+m_f^2) e^{-NV(\lambda_k^2)} 
\Delta(\{\lambda_i\})^2
\end{equation}
the $k$-point correlation function is given by
\begin{equation} \label{kpoint}
\rho_k(\lambda_1,\ldots,\lambda_k)~=~
\intR \prod_{i=k+1}^N \sd \lambda_i \rho_N(\lambda_1,\ldots,\lambda_N) ~.
\end{equation}
The orthodox way to calculate these correlators would be to f\/ind orthogonal polynomials 
$\{\mathcal{P}_n\}$ with respect to the weight function
\begin{equation} \label{UEmassive} 
w(\lambda) ~=~ \lambda \prod_{f=1}^{\alpha} (\lambda^2+m_f^2) e^{-NV(\lambda^2)} ~,
\end{equation}
introduce the kernel
\begin{equation}
K_N(\lambda,\lambda') ~=~ \sqrt{w(\lambda) w(\lambda')}
\sum_{k=0}^{N-1} \frac{\mathcal{P}_k(\lambda)\mathcal{P}_k(\lambda')}{h_k}  
\end{equation}
where $h_k$ is the norm of $\mathcal{P}_k$, and then use the expression
\begin{equation}
\rho_k(\lambda_1,\ldots,\lambda_k)~=~\frac{(N-k)!}{k!}
\det_{1 \le a,b \le k} K_N(\lambda_a,\lambda_b)
\end{equation}
to determine the $k$-point correlation functions. This method cannot be used here since 
$w(\lambda)$ is odd and the integration interval is even, and thus orthogonal polynomials are 
not def\/ined. 

Inspired by \cite{VZ} we will now introduce a method based on what we will call 
pseudo-orthogonal polynomials that can be applied to any odd measure on an even interval as long
as $N$ is even. Consider for simplicity the massless case where 
$w(\lambda)=\lambda^{2\alpha+1}e^{-NV(\lambda^2)}$. For the chiral unitary ensemble (chUE, with
$\beta=2$) the weight function for $N_f^{ch}$ massless fermions in the sector of topological 
charge $\nu$ is
\begin{equation}
w_{ch}(\lambda)~=~\lambda^{N_f^{ch}+\nu}e^{-NV(\lambda)} \ , \ \lambda \in \mathbb{R}_{+} 
\end{equation}
and we denote the associated orthogonal polynomials $\{P_n(\lambda)\}$ with norm $h_n$. Now put 
$N_f^{ch}=\alpha$, $\nu=\frac{1}{2}$, and def\/ine the pseudo-orthogonal polynomials by
\begin{eqnarray}
Q_{2n}(\lambda) &\equiv& (1+\lambda)P_n(\lambda^2) \label{def1} \\
Q_{2n+1}(\lambda) &\equiv& (1-\lambda)P_n(\lambda^2) \label{def2} ~.
\end{eqnarray}
These new polynomials are pseudo-orthogonal with respect to the weight function 
$w(\lambda)=\lambda^{2\alpha+1}e^{-NV(\lambda^2)}$ on the real line in the sense that
\begin{eqnarray}
\bra{$2k$} \cket{$2n$} &=& 2 h_n \delta_{kn} \ \ \ \!   \equiv r_{2n} \delta_{kn} \\
\bra{$2k+1$} \cket{$2n+1$}&=& -2 h_n \delta_{kn} \equiv r_{2n+1} \delta_{kn} \\
\bra{$2k$} \cket{$2n+1$}&=& 0
\end{eqnarray}
where the brackets indicate integration. The proof is straightforward. The chUE-measure for 
$N_f^{ch}$ massive f\/lavors is
\begin{equation} 
w_{ch}(\lambda)~=~ \prod_{f=1}^{N_f^{ch}}(\lambda+m_f^2) e^{-NV(\lambda)} \ , \ 
\lambda \in \mathbb{R}_{+} ~.
\end{equation}
The measure is valid in the sector of vanishing topological charge. In the sector of charge 
$\nu \not= 0$ we obtain the correct measure by adding $\abs{\nu}$ massive f\/lavors and 
then letting the masses vanish. We see that in the sector where $\nu=\frac{1}{2}$ and 
$N_f^{ch}=\alpha$ the measure is
\begin{equation} \label{chUEmassive}
w_{ch}(\lambda)~=~ \sqrt{\lambda} \prod_{f=1}^{\alpha}(\lambda+m_f^2) e^{-NV(\lambda)} \ , \ 
\lambda \in \mathbb{R}_{+} ~.
\end{equation} 
Simple calculations shows that the pseudo-orthogonal polynomials with respect to 
(\ref{UEmassive}) are given by the polynomials orthogonal with respect to (\ref{chUEmassive}) by
the def\/initions (\ref{def1}) and (\ref{def2}). Because the QCD$_3$ masses are pairwise assigned
with opposite sign we may identify the $f$'th chiral mass with the positive mass of the $f$'th 
QCD$_3$ fermion. Notice that in the matrix representation of chUE half-integer values of $\nu$ 
are not permitted since the matrices integrated over are $N \times (N+\abs{\nu})$ \cite{S-V}. 
In the eigenvalue representation, however, we can without problems make an analytic continuation
such that $\nu=\frac{1}{2}$ makes sense, at least from a formal point of view.

We are now ready to calculate the microscopic spectral correlators of odd-f\/lavored 
QCD$_3$, \ie,
the correlation functions taken in the limit $N \to \infty$, $\lambda \to 0$, $m_f \to 0$ where
\begin{equation}
\zeta ~\equiv~ \pi \rho(0) N \lambda  \ , \  \zeta' ~\equiv~ \pi \rho(0) N \lambda'  \ , \ 
\mu_f ~\equiv~  \pi \rho(0) N m_f  
\end{equation}
are kept constant. We always assume that $N$ is even. $\Delta(\{\lambda_k\})$ def\/ined in 
(\ref{Vandermonde}) can be expressed as the determinant of an $N \times N$ matrix:
\begin{equation} \label{VdM}
\Delta(\{\lambda_k\}) ~=~ \det \left( \begin{array}{cccc} 
1           &         1   & \cdots &         1 \\
\lambda_1   & \lambda_2   & \cdots & \lambda_N  \\
\lambda_1^2 & \lambda_2^2 & \cdots & \lambda_N^2 \\
\vdots & \vdots & & \vdots \\
\lambda_1^{N-1} & \lambda_2^{N-1} & \cdots & \lambda_N^{N-1} 
\end{array} \right)  ~.
\end{equation}
Consider the $N \times N$ matrix
\begin{equation}
\mathcal{C}(\{\lambda_k\})~=~\left( \begin{array}{cccc} 
1+\lambda_1 & 1+\lambda_2 & \cdots & 1+\lambda_N \\
1-\lambda_1 & 1-\lambda_2 & \cdots & 1-\lambda_N  \\
\lambda_1^2+\lambda_1^3 & \lambda_2^2 + \lambda_2^3& \cdots & 
\lambda_N^2+\lambda_N^3 \\
\lambda_1^2-\lambda_1^3 & \lambda_2^2-\lambda_2^3& \cdots & 
\lambda_N^2-\lambda_N^3 \\
\vdots & \vdots & & \vdots \\
\lambda_1^{N-2}+\lambda_1^{N-1} & \lambda_2^{N-2}+\lambda_2^{N-1} & \cdots & 
\lambda_N^{N-2}+\lambda_N^{N-1} \\
\lambda_1^{N-2}-\lambda_1^{N-1} & \lambda_2^{N-2}-\lambda_2^{N-1} & \cdots & 
\lambda_N^{N-2}-\lambda_N^{N-1} \\ 
\end{array} \right) ~.
\end{equation}
The crucial observation is that $\det \mathcal{C}(\{\lambda_k\})$ is related to 
$\Delta(\{\lambda_k\})$. Performing on $\mathcal{C}$ the row operations
\begin{equation}
\tx{row}(2k-1) ~\to~ \tx{row}(2k-1)+\tx{row}(2k) \ , \ k ~=~ 1,\ldots,N/2 
\end{equation}
followed by
\begin{equation}
\tx{row}(2k) ~\to~ \tx{row}(2k)-\frac{1}{2}\tx{row}(2k-1) \ , 
 \ k ~=~ 1,\ldots,N/2
\end{equation}
we end up with
\begin{equation}
\tilde{\mathcal{C}}(\{\lambda_k\})~=~\left( \begin{array}{cccc} 
2 & 2 & \cdots & 2 \\
-\frac{1}{2}\lambda_1 & -\frac{1}{2}\lambda_2 & \cdots 
          & -\frac{1}{2}\lambda_N \\
\vdots & \vdots & & \vdots \\
2\lambda_1^{N-2} & 2\lambda_2^{N-2} & \cdots & 2\lambda_N^{N-2} \\
-\frac{1}{2}\lambda_1^{N-1} & -\frac{1}{2}\lambda_2^{N-1} & \cdots & 
-\frac{1}{2}\lambda_N^{N-1} \\ 
\end{array} \right) ~,
\end{equation}
where we have discarded a factor coming from the row operations. The determinant of an 
$N \times N$ matrix $M$ is def\/ined as
\begin{equation}
\tx{det}M ~=~ \sum_{\pi \in S_N}(-)^{\pi} M_{1 \pi(1)} \cdots M_{N \pi(N)} 
\end{equation}
from which it follows that 
$\det \mathcal{C}(\{\lambda_k\})=(-)^{N/2} \Delta(\{\lambda_k\})$, or
\begin{equation}
(-)^{N/2}\det \mathcal{C}(\{\lambda_k\}) ~=~ (-)^N \Delta(\{\lambda_k\}) 
~=~ \Delta(\{\lambda_k\}) ~.
\end{equation}
Now notice that
\begin{equation} \label{VdMlike}
\mathcal{C}(\{\lambda_k\})~=~\left( \begin{array}{llll} 
(1+\lambda_1)1 & (1+\lambda_2)1 & \cdots & (1+\lambda_N)1 \\
(1-\lambda_1)1 & (1-\lambda_2)1 & \cdots & (1-\lambda_N)1  \\
(1+\lambda_1)\lambda_1^2 & (1+\lambda_2)\lambda_2^2 & \cdots & 
(1+\lambda_1)\lambda_N^2 \\
(1-\lambda_1)\lambda_1^2 & (1-\lambda_2)\lambda_2^2 & \cdots & 
(1-\lambda_1)\lambda_N^2 \\
\ \ \ \ \ \ \vdots & \ \ \ \ \ \ \vdots & & \ \ \ \ \ \ \vdots \\
(1+\lambda_1)\lambda_1^{N-2} & (1+\lambda_1)\lambda_2^{N-2} & \cdots & 
(1+\lambda_1)\lambda_N^{N-2} \\
(1-\lambda_1)\lambda_1^{N-2} & (1-\lambda_1)\lambda_2^{N-2} & \cdots & 
(1-\lambda_1)\lambda_N^{N-2}  \\ 
\end{array} \right) ~.
\end{equation}
Looking at (\ref{VdMlike}) we see that we are in a position where we can substitute the entries
in $\mathcal{C}(\{\lambda_k\})$ with the pseudo-orthogonal polynomials (properly normalized). 
Because of pseudo-orthogonality we may then write, as in the case of orthogonal polynomials,
\begin{equation} 
K_N(\lambda,\lambda')~=~ \sqrt{\lambda \lambda'} \prod_{f=1}^{\alpha} \sqrt{\lambda^2+m_f^2}
\sqrt{{\lambda'}^2+m_f^2}  e^{-\frac{N}{2} (V(\lambda^2)+V({\lambda'}^2))} 
\frac{1}{N} \sum_{j=0}^{N-1} \frac{Q_j(\lambda) Q_j(\lambda')}{r_j} ~,
\end{equation}
where the $Q_j$'s are pseudo-orthogonal. With $n \equiv \frac{N}{2}$ we obtain
\begin{eqnarray}
\sum_{k=0}^{N-1} \frac{Q_k(\lambda) Q_k (\lambda')}{r_k} &=& 
\sum_{i=0}^{n-1} \frac{1}{r_{2i}}(1+\lambda)(1+\lambda') P_{i}(\lambda^2)P_{i}({\lambda'}^2) +
\sum_{j=0}^{n-1} \frac{1}{r_{2j+1}}(1-\lambda)(1-\lambda') P_{j}(\lambda^2)P_{j}({\lambda'}^2) 
\nonumber \\
&=& 
\sum_{i=0}^{n-1} \frac{1}{r_{2i}}(1+\lambda)(1+\lambda') P_{i}(\lambda^2)P_{i}({\lambda'}^2) 
- \sum_{j=0}^{n-1} \frac{1}{r_{2j}}(1-\lambda)(1-\lambda') P_{j}(\lambda^2)P_{j}({\lambda'}^2) 
\nonumber \\
&=& (\lambda+\lambda') \sum_{k=0}^{n-1} \frac{P_{k}(\lambda^2) P_{k} ({\lambda'}^2)}{h_k} ~,   
\end{eqnarray} 
We have discarded irrelevant overall factors. The introduction of 
pseudo-orthogonal polynomials have led us to an expression for the kernel that involves 
orthogonal polynomials for which the microscopic limit is universal 
\cite{GA-PHD-UM-SN,PHD-NISH}:
\begin{equation} \label{QCD3kernel}
K_N(\lambda,\lambda') ~=~  
(\lambda+\lambda')\sqrt{\lambda \lambda'} \prod_{f=1}^{\alpha} \sqrt{\lambda^2+m_f^2}
\sqrt{{\lambda'}^2+m_f^2} e^{-\frac{N}{2} (V(\lambda^2)+V({\lambda'}^2))} \frac{1}{N} 
\sum_{j=0}^{n-1} \frac{P_j(\lambda^2) P_j({\lambda'}^2)}{h_j} ~.
\end{equation}
By the Christof\/fel-Darboux formula the sum over the polynomials is proportional to
$[P_{n-1}(\lambda) P_{n}(\lambda')-P_n(\lambda) P_{n-1}(\lambda')]/(\lambda^2-{\lambda'}^2)$
and since $n=\frac{N}{2}$ the $N \to \infty$ limit corresponds to the $n \to \infty$ limit. Now
compare (\ref{QCD3kernel}) with the expression for the chUE kernel \cite{PHD-NISH}
\begin{eqnarray} \label{chUEkernel}
K_{\tx{chUE},N}(z_1,z_2)&=&\sqrt{\abs{z_1 z_2}} 
\prod_{f=1}^{N_f^{ch}} \sqrt{z_1^2+m_f^2} \sqrt{z_2^2+m_f^2}e^{-\frac{N}{2}(V(z_1^2)+V(z_2^2)}  
\nonumber \\
&&\times~ \frac{P_{N-1}(z_1^2) P_N(z_2^2) - P_N(z_1^2) P_{N-1} (z_2^2)}{z^2_1-z^2_2}
\end{eqnarray}
in the sector of zero topological charge. To go to the sector of charge $\nu$ we can, by 
f\/lavor-topology duality, take the $\nu=0$ kernel with $N_f^{ch}+\abs{\nu}$ massive 
f\/lavors and 
then let $\abs{\nu}$ masses vanish. The sector with $\nu=\frac{1}{2}$ is then by continuation 
given by (\ref{chUEkernel}) multiplied by a factor $\sqrt{z_1 z_2}$. We arrive at the following
master formula for odd-f\/lavored QCD$_3$:
\begin{equation} \label{kernelrelation}
K_{s}^{(N_{f}=2\alpha+1)}(0,\mu_1,\ldots,\mu_{\alpha},-\mu_1,\ldots,-\mu_{\alpha},\zeta_1,\zeta_2)
~=~ \frac{\zeta_1+\zeta_2}{\sqrt{\zeta_1\zeta_2}}
K_{{\mbox{\rm chUE}},s}^{(N_{f}^{ch}=\alpha,\nu=1/2)}
(\mu_1,\ldots,\mu_{\alpha},\zeta_1,\zeta_2) 
~. 
\end{equation}
The formula implies that all properties of the chUE that are universal similarly are universal in
odd-f\/lavored QCD$_3$ and especially that all microscopic correlators are universal. We see 
that in the case of the microscopic spectral density, where we have to evaluate 
$K_s(\zeta,\zeta)$, the odd-f\/lavored QCD$_3$ kernel and the chUE kernel agree up to a 
proportionality constant. They do not, however, agree for higher-order correlators. The chUE 
kernel is given by \cite{PHD-GA, PHD-NISH} 
\begin{equation} \label{chiralkernel}
K_{\tx{chUE},s}(\mu_1,\ldots,\mu_{\alpha},\zeta_1,\zeta_2) ~=~ 
C_2 \frac{(-)^{\nu+[N_f^{ch}/2]+1} \sqrt{\zeta_1\zeta_2}}{(\zeta_1^2-\zeta_2^2) 
\prod_f 
\sqrt{(\zeta_1^2+\mu_f^2)(\zeta_2^2+\mu_f^2)}} \frac{\det \mathcal{B}}{\det \mathcal{A}} ~,
\end{equation}
where $C_2$ is a normalization constant, and where $\mathcal{B}$ is an 
$(N_f^{ch}+2) \times (N_f^{ch}+2)$ matrix and $\mathcal{A}$ is an $N_f^{ch} \times N_f^{ch}$ 
matrix def\/ined as
\begin{eqnarray}
\mathcal{A}_{ij} &=& \mu_i^{j-1} I_{\nu+j-1} (\mu_i)\\
\mathcal{B}_{ij} &=& (\zeta_i)^{j-1} J_{\nu+j-1}(\zeta_i) \ \quad  \quad \quad , \ i=1,2 \\
\mathcal{B}_{ij} &=& (-\mu_{i-2})^{j-1} I_{\nu+j-1} (\mu_{i-2}) \ , \ i=3,\ldots,N_f^{ch}+2  ~.
\end{eqnarray}
In QCD$_3$ the hitherto only known microscopic spectral correlator is the massless microscopic 
spectral density \cite{VZ}
\begin{equation}
\rho_s^{(2\alpha+1)}(\zeta) ~=~ \frac{\zeta}{2} 
\left[ J_{\alpha+\frac{1}{2}}^2(\zeta) - 
J_{\alpha+\frac{3}{2}}(\zeta) J_{\alpha-\frac{1}{2}}(\zeta) \right]
\end{equation}
which coincides with the massless chUE microscopic spectral density
\begin{equation}
\rho_{\tx{chUE},s}^{(N_f^{ch},\nu)}(\zeta) ~=~ \frac{\zeta}{2}
\left[ J_{N_f^{ch}+\nu}^2(\zeta) - J_{N_f^{ch}+\nu+1}(\zeta) J_{N_f^{ch}+\nu-1}(\zeta) \right]
\end{equation}
precisely when $N_f^{ch}=\alpha$ and $\nu=\frac{1}{2}$. 

\begin{figure} 
\input epsf
\epsfxsize=10cm
\epsfysize=7cm
\centerline{ {\epsffile{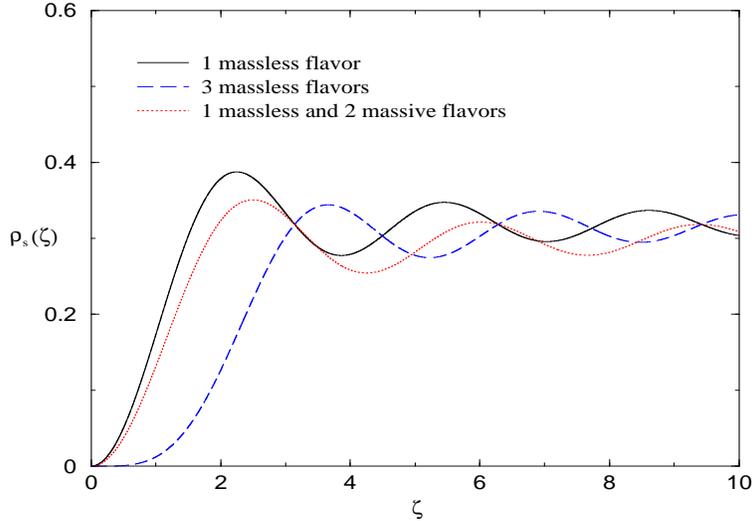} } }
\caption{The microscopic spectral density in three dif\/ferent cases: 
$\rho^{(1)}_s(0,\zeta)$ 
(solid curve), $\rho^{(3)}_s(0,0,0,\zeta)$ (dashed curve), and $\rho^{(3)}_s(0,\mu,-\mu,\zeta)$
with $\mu=10$ (dotted curve). In the $\mu \to \infty$ limit the dotted curve will approach the 
solid curve, cf. the decoupling relations (\ref{decoupl}). In the $\mu \to 0$ limit the dotted 
curve will converge to the dashed curve as the massive spectral density becomes massless.}
\label{densities}
\end{figure}

From (\ref{kernelrelation}) it directly follows that the microscopic spectral density obeys a
series of decoupling relations \cite{PHD-NISH}:
\begin{equation} \label{decoupl}
\lim_{\mu_k \to \infty} 
\rho^{(2k+1)}_s (0,\mu_1,\ldots,\mu_k,\zeta) ~=~ 
\rho^{(2k-1)}_s (0,\mu_1,\ldots,\mu_{k-1},\zeta) \ , \ k=1,\ldots,\alpha ~.
\end{equation}
These relations tell us that when we make two quarks with equal masses (up to a sign) 
inf\/initely heavy the spectral density of the new system becomes that of the system 
consisting 
of the remaining quarks. This is illustrated in f\/igure \ref{densities} which shows the 
massless
microscopic spectral density (msd) for one and three massless f\/lavors, and for one 
massless and
two massive f\/lavors. The massive msd is calculated using the master formula 
(\ref{kernelrelation}). When the masses vanish the massive msd will become that of three
massless fermions while the act of making the masses inf\/initely heavy will make the 
massive
msd become that of one massless f\/lavor. 

\section{The Kernel in Terms of F\/inite-volume Partition Functions} 
\label{sec: finite-vol}
The newly acquired universality of the microscopic spectral correlators will now allow us to
express the odd-f\/lavored RMT kernel for even $N$ in terms of the f\/ield theory 
f\/inite-volume partition functions of QCD$_3$, as it was done in \cite{PHD-GA,PHD} for the 
other ensembles. Having done this we go on to simplify the expression for the f\/inite-volume 
partition functions considerably when we take the conditions for parity invariance into account.

The main ingredient in the proof is a result from \cite{Zinn-Justin} that in our case says that
\begin{eqnarray} \label{kernel}
K^{(N_f)}_N(\lambda,\lambda')~=~\frac{1}{\Zt_N} 
e^{-\frac{N}{2}[V(\lambda^2)+V({\lambda'}^2)]}
\sqrt{\lambda \lambda'} \prod_{f=1}^{\alpha} \sqrt{(\lambda^2+m^2_f)({\lambda'}^2+m^2_f)} 
\nonumber \\
\times \int \tx{d}^{{(N-1)}^2} M e^{-N\tx{tr}V(M^2)} \det(M) \prod_{f=1}^{\alpha} \det(M^2+m_f^2)
\det(\lambda-M) \det(\lambda'-M) ~.
\end{eqnarray}
The integral appearing in the second line is the RMT partition function in the case where the
square hermitian matrices are of odd dimensions, and with two extra imaginary masses $i\lambda$
and $i\lambda'$. The proof goes as in \cite{Zinn-Justin}:
\begin{equation}
\Zt_{N-1}(\mm,i\lambda,i\lambda') ~=~ \int \sd^{(N-1)^2} M \det(\lambda-M) \det(\lambda'-M) W(M)
\end{equation}
where $W(M)=\Zt_N^{-1}e^{-N\tx{tr}V(M^2)} \det(M) \prod_{f=1}^{\alpha} \det(M^2+m_f^2)$ is the 
RMT measure. Rewriting in terms of eigenvalue variables we get
\begin{equation}
\Zt_{N-1}(\mm,i\lambda,i\lambda') ~=~ \intR \prod_{k=1}^{N-1} \sd \lambda_k (\lambda-\lambda_k)
(\lambda'-\lambda_k) \Delta^2(\{\lambda_i\}_{i=1}^{N-1}) w(\lambda_k)
\end{equation}
where $w(\lambda_k)$ is given by (\ref{UEmassive}). Now absorb the factors of 
$(\lambda-\lambda_k)$ and $(\lambda'-\lambda_k)$ into the Vandermonde determinants:
\begin{eqnarray}
\Delta_{\lambda} &\equiv& \Delta(\lambda,\{\lambda_i\}_{i=1}^{N-1}) \\
\Delta_{\lambda'}     &\equiv& \Delta(\lambda',\{\lambda_i\}_{i=1}^{N-1}) ~.
\end{eqnarray}
These new Vandermonde determinants are $N \times N$, $N$ even, and thus we have the method of
pseudo-orthogonal polynomials at our disposal.
\begin{equation}
\Zt_{N-1}(\mm,i\lambda,i\lambda') ~=~ 
\intR \prod_{k=1}^{N-1} \sd \lambda_k w(\lambda_k) \Delta_{\lambda} \Delta_{\lambda'} 
          ~=~ \sum_{k=0}^{N-1}\frac{Q_k(\lambda)Q_k(\lambda')}{r_k} ~,
\end{equation}
where the $Q_k$ are the pseudo-orthogonal polynomials corresponding to $w(\lambda_k)$. Note that
the proof does not carry through when $N$ is odd. Now, because of universality the RMT 
partition function $\Zt$ is equal (up to a constant) to the f\/ield theory 
f\/inite-volume 
partition function $\Z$ and thus
\begin{equation} \label{2oddkernel}
K^{(N_f)}_s (\eta=+1;\mm,\zeta,\zeta') = 
C^{(N_f)} \sqrt{\zeta \zeta'} \prod_{f=1}^{\alpha} 
\sqrt{(\zeta^2+\mu_f^2)({\zeta'}^2+\mu_f^2)}   
\frac{\Z^{(N_f+2)} (\eta=-1;\mm, i\zeta, i\zeta')}{\Z^{(N_f)}(\eta=+1;\mm)} ~.
\end{equation}
In formula (\ref{2oddkernel}) $\mm$ now has the microscopic masses as entries, and the 
partition functions have been labelled with $N$ and $N-1$ implicitly through 
$\eta \equiv (-)^N$. Notice that we have already taken the double-microscopic limit - the label
is important because the partition functions depend on $N$ in a crucial manner. We have 
\cite{PHD-GA}
\begin{equation} \label{Z_N}
\Z^{(N_f)}(\eta; \mu,\{\mu_k\})~=~\frac{1}{\Delta(\mathcal{M})}
\left[  \det\mathbf{D}(\mu,\{\mu_k\})
+(-)^{N+\alpha} \det\mathbf{D}(-\mu,\{-\mu_k\}) \right] ~. 
\end{equation}
In the expression $\Delta(\mathcal{M})$ is a Vandermonde determinant of the masses and 
$\mathbf{D}(\mu,\{\mu_k\})$ is an $N_f \times N_f$ matrix of the form ($\mu_0 \equiv \mu$, 
$\mu_{\alpha+i}=-\mu_i \tx{ for } i=1,\ldots,\alpha)$ 
\begin{eqnarray}
\mathbf{D}_{ij} &=& \mu_{i-1}^{j-1} e^{\mu_{i-1}} \ \ \ \ \ \ \ \ \ \ \ \ \ \ \ \ , 
\ j~=~1,\ldots,\alpha+1 \label{D1} \\
\mathbf{D}_{ij} &=& (-\mu_{i-1})^{j-\alpha-2} e^{-\mu_{i-1}} \label{D2}
\ \ , \  j~=~\alpha+2,\ldots,2\alpha+1 ~.
\end{eqnarray}
Formula (\ref{Z_N}) can be simplif\/ied considerably when $\mu=0$. Take the matrix 
$\mathbf{D}(-\mu,\{-\mu_k\})$ and interchange
\begin{equation} 
\tx{row } k ~\leftrightarrow~ \tx{row } \alpha+k ~,~k=2,\ldots,\alpha+1 
\end{equation}
such that the rows $\alpha+2$ to $2\alpha+1$ become positioned right below the $\mu$-dependent 
row. Because the interchange of two rows amounts to an overall sign change of the determinant we
see that
\begin{equation}
\mathbf{D}(-\mu=0,\{-\mu_k\}) ~=~ (-)^{\alpha}\det\mathbf{D}(\mu=0,\{\mu_k\})
\end{equation}
because of the $\pm$ symmetry of the masses and the block structure of (\ref{D1})-(\ref{D2}).
Thus
\begin{eqnarray}
\Z^{(N_f)}(\eta; \mu=0,\{ \mu_k \}) &=& 
\frac{1}{\Delta(\mathcal{M})}
\left[ \det\mathbf{D}(\mu=0,\{\mu_k\})  + 
(-)^{N+2\alpha} \det\mathbf{D}(\mu=0,\{\mu_k\}) \right] \nonumber \\
&=& \left\{  \begin{array}{ll}
2 \det\mathbf{D}(\mu=0,\{\mu_k\}) & \tx{if $N$ is even} \\
0                               & \tx{if $N$ is odd} ~.
\end{array} \right. 
\end{eqnarray}
Besides the simplif\/ication this expression tells us that we should not look upon the odd-$N$
partition function as def\/ining a theory in itself, but only as a function that enters in the
expressions of the even-$N$ theory. 

Note that when $\nu=\frac{1}{2}$ the chUE kernel (\ref{chiralkernel}) becomes complex and thus
the normalization coef\/f\/icient
\begin{equation}
C_2 ~=~ (-)^{\nu+[N^{ch}_f/2]}
\end{equation}
becomes complex also. We have explicitly checked that (\ref{2oddkernel}) gives the expected 
results in a few cases, and in each case the normalization constants had to be chosen complex. 
F\/inally, we mention that the decoupling relations (\ref{decoupl}) can be explicitly 
derived from (\ref{2oddkernel}). 

There are two major points to take away from this section. F\/irst, the fact that for even $N$
we are able to rewrite the RMT kernel in terms of f\/inite-volume partition functions while the
proof cannot be applied to the case of odd $N$. Second, the fact the f\/inite-volume partition
function for odd $N$ vanishes when the only input is the mass assignment that guarantees 
parity invariance, a crucial point when the RMT is constructed. From this we conclude that when
extracting spectral information about odd-f\/lavored QCD$_3$ from RMT we should look upon the 
even-$N$ theory as def\/ining the theory while the odd-$N$ f\/inite-volume partition function 
should be looked upon merely as a function that enters in the expression for the even-$N$ kernel.

\section{Summary}
We have considered the RMT formulation of QCD$_3$ with an odd number of f\/lavors in the 
mesoscopic scaling region and we have calculated the microscopic kernel from which all 
microscopic spectral correlation functions can be found. We have shown the universality of the 
kernel by the discovery of a deep connection between the kernel of odd-f\/lavored QCD$_3$ with 
$N_f=2\alpha+1$ flavors and QCD$_4$ ($\beta=2$) with $N_f^{ch}=\alpha$ f\/lavors in the sector 
of topological charge $\nu=\frac{1}{2}$, brought about by the introduction of pseudo-orthogonal
polynomials. The result has the consequence that all properties of the four-dimensional theory 
that are universal similarly are universal in three dimensions. We end by noting that 
odd-f\/lavored QCD$3$ now f\/inally has caught up with its older and more mature sisters QCD$_4$
and even-f\/lavored QCD$_3$ as far as universality is concerned. 

\vspace{0.3cm}
\noindent
\textsf{Acknowledgements:} P. H. Damgaard, A. Andersen, and K. Splittorf\/f are thanked for 
discussions and critical reading of the manuscript. 

\end{document}